\documentclass[sigconf]{acmart}
\AtBeginDocument{%
  \providecommand\BibTeX{{%
    \normalfont B\kern-0.5em{\scshape i\kern-0.25em b}\kern-0.8em\TeX}}}
\usepackage{multirow}
\usepackage{graphicx}
\usepackage{array}
\usepackage{ragged2e}
\copyrightyear{2024} 
\acmYear{2024} 
\setcopyright{acmlicensed}\acmConference[CHI '24]{Proceedings of the CHI Conference on Human Factors in Computing Systems}{May 11--16, 2024}{Honolulu, HI, USA}
\acmBooktitle{Proceedings of the CHI Conference on Human Factors in Computing Systems (CHI '24), May 11--16, 2024, Honolulu, HI, USA}
\acmDOI{10.1145/3613904.3642175}
\acmISBN{979-8-4007-0330-0/24/05}

\begin{document}

\title[Circle Back Next Week]{Circle Back Next Week: The Effect of Meeting-Free Weeks on Distributed Workers' Unstructured Time and 
Attention Negotiation}

\author{Sharon Ferguson}
\email{sharon.ferguson@mail.utoronto.ca}
\orcid{0000-0002-2091-3435}
\affiliation{%
  \institution{Mechanical and Industrial Engineering, \\University of Toronto}
  \streetaddress{55 St. George St}
  \city{Toronto}
  \state{Ontario}
  \country{Canada}
  \postcode{M5S 0C9}
}

\author{Michael Massimi}
\email{mmassimi@slack-corp.com}
\orcid{0009-0007-8949-7269}
\affiliation{%
  \institution{Workforce Lab, Slack}
  \streetaddress{500 Howard St}
  \city{San Francisco}
  \state{California}
  \postcode{94105}
  \country{United States}}







\renewcommand{\shortauthors}{Ferguson et al.}

\begin{abstract}

While distributed workers rely on scheduled meetings for coordination and collaboration, these meetings can also challenge their
ability to focus. Protecting worker focus has been addressed from a technical perspective, but companies are now attempting
organizational interventions, such as meeting-free weeks. Recognizing distributed collaboration as a sociotechnical challenge, we
first present an interview study with distributed workers participating in meeting-free weeks at an enterprise software company. We
identify three orientations workers exhibit during these weeks: Focus, Collaborative, and Time-Bound, each with varying levels and use of unstructured time. These different orientations result in challenges in attention negotiation, which may be suited for technical
interventions. This motivated a follow-up study investigating attention negotiation and the compensating mechanisms workers developed
during meeting-free weeks. Our framework identified tensions between the attention-getting and attention-delegation strategies. We
extend past work to show how workers adapt their virtual collaboration mechanisms in response to organizational interventions.

\end{abstract}




\received{14 September 2023}
\received[revised]{12 December 2023}
\received[accepted]{18 January 2024}

\maketitle

\section{Introduction}

As of October 2023, 12\% of full-time American employees were working remotely, and 30\% were hybrid \cite{wfhreport}, with these numbers expected to increase \cite{BenWigert_2023}. For decades, Human-Computer Interaction (HCI) researchers have studied distributed teams, identifying current challenges \cite{10.1145/3462204.3481780, 10.1145/587078.587110, 10.1145/3411763.3451793}, and designing technical solutions to overcome them \cite{10.1145/3584931.3607502, 10.1145/1031607.1031678, 10.1145/2141512.2141535}. One well-documented challenge in distributed collaboration is a loss of awareness of colleagues' work \cite{ferguson2022couldn, 10.1145/3544548.3580989}. It is well documented that in response, distributed teams add more meetings \cite{yang2022effects, ferguson2022couldn, 10.1145/3544548.3580989}. While meeting overload is a problem that has been reported for years \cite{luong2005meetings}, distributed working practices are bringing this phenomenon to the forefront.

Knowledge workers report that meetings can take up more than 85\% of their day \cite{Chen_2020}, 70\% of meetings keep them from getting their work done, and 92\% of workers believe meetings are costly and unproductive \cite{laker2022dear}. This is a frustrating experience that affects workers' well-being \cite{luong2005meetings} and productivity \cite{gibbs2021work}. Meetings are believed to hurt productivity by interrupting individuals and preventing them from focusing on deep work \cite{Mark2015}. The increased flexibility in distributed workers' schedules may exacerbate this problem --- leading to more fragmented schedules \cite{team_2023}.

Improving workers' focus is not a new challenge for the HCI community. Researchers have developed systems that automatically schedule focus time in workers' calendars \cite{saha2023focus} or block distracting websites when one is susceptible to interruptions \cite{tseng2019overcoming}. However, protecting workers' focus time is both a technical and an organizational problem. One can be interrupted by incoming requests from teammates (technology-mediated or not) \cite{mark2005no}, scheduled organizational events (e.g., meetings) \cite{Brown_2023}, notifications from collaboration tools \cite{karr2010more}, or even themselves \cite{mark2005no}. 

One organizational intervention that companies are experimenting with to improve focus time is designating organization-wide meeting-free time \cite{laker2022surprising}. 
For example, Shopify recently declared ``meeting bankruptcy'' and encouraged workers to think critically before scheduling meetings \cite{Boyle_2023}.
Organizations believe cancelling meetings will provide more unstructured time that workers will use towards independent focus \cite{gibbs2021work}. In fact, these emerging programs are not well understood, with a number of open questions: how is unstructured time actually used? What are the benefits and drawbacks of this intervention? And ultimately, does meeting-free time improve perceived focus? With a better understanding of how these organizational interventions address the problem of worker focus, we can identify gaps suited for technical interventions. 

We first contribute an interview study (n=17) of a distributed enterprise software development organization that implemented meeting-free weeks. In this work, meeting-free weeks refer to the cancellation of \textit{scheduled meetings}, defined as pre-scheduled synchronous collaboration conducted over video/audio calls with two or more people, as well as the encouragement to avoid adding new meetings. We aimed to develop an understanding of how knowledge workers spent their newfound unstructured time, whether they perceived an increase in focus time, and the additional perceived benefits and drawbacks. Using semi-structured interviews with 17 workers, we identified three ways workers used unstructured time during meeting-free weeks:
\begin{itemize}
    \item The Focus Orientation: A chance for independent focus in service of productivity or discovery.
    \item The Collaborative Orientation: An opportunity for synchronous collaboration that workers would not ordinarily have time for, such as ad-hoc meetings that help others push their work forward. 
    \item The Time-Bound Orientation: Little change because the demands of the worker's role do not change in the absence of meetings.
\end{itemize}
A major finding from the interviews was that because workers 
used this unstructured time differently and could not fall back on their default attention-getting method of scheduled meetings, 
they found it challenging to negotiate the attention of their teammates. They had to rely on new virtual coordination mechanisms to get others' attention, which were not always successful.

As distributed workers have no choice but to rely on technology to get the attention of their teammates, it may be possible for tailored design solutions to address the existing tensions.
To better understand this challenge, we conducted a follow-up interview study to see how workers currently negotiate attention in the absence of meetings.
Our subsequent research questions were: Without meetings, how do distributed workers negotiate attention? What compensating mechanisms are developed? What new challenges arise as a result? Using semi-structured interviews, we asked 11 workers to guide us through a scenario where they attempted to get others' attention and had to decide what to pay attention to themselves. 

From this evidence, we developed a framework describing virtual attention negotiation using compensating mechanisms, or mechanisms other than scheduled meetings, to support organizations' goals of reducing the overall number of meetings. We identified tensions that arise when the strategies workers use to \emph{get} attention do not align with the strategies others use to \emph{delegate} their attention, resulting in ineffective attention negotiation that makes workers think they need a synchronous meeting.
Researchers have independently studied how to build awareness of others' attentional states \cite{liu2022exploratory, 10.1145/3406865.3418308}, and how to protect one's own attention \cite{tseng2019overcoming}. By identifying the strategies used on both sides of this negotiation, we develop design considerations for systems that support both parties in this collective negotiation, without resorting to synchronous meetings.

We approach distributed workplace collaboration \cite{10.1145/3411764.3445243, park2023towards, 10.1145/3544548.3581141, 10.1145/3544548.3580989} holistically, as a system that requires both organizational and technical considerations. We build on the body of HCI literature solving the problem of competing demands for workers' focus \cite{blank2020emotional, Brumby2019, mark2018effects, mark2017blocking, mark2016email, mark2015focused, mark2014bored, dabbish2011keep, mark2008cost, 10.1145/633292.633351, czerwinski2004diary, gross2023literature} by providing an understanding of how and when meeting-free weeks can help. Most work requires a mix of both independent and collaborative activities, and how one person uses their time can affect other individuals' and groups' work \cite{perlow1999}. We show that distributed workers develop compensating mechanisms, adapting the way that they use their collaborative tools, in response to the intervention of meeting-free weeks. 
These compensating mechanisms are used for attention negotiation, a process that involves the interplay of the attention-getter, the attention-target, and the technology between them. Extending the work on awareness in remote collaboration \cite{liu2022exploratory, 10.1145/3406865.3418308}, we show how current virtual collaboration tools do not promote awareness of the nuanced contextual details used in attention negotiation, such as personal preferences, team norms, and personal platform settings. We provide organizational and technical implications for this challenging sociotechnical problem.

\section{Background}

%
Globalization, COVID-19, and technological advancements have increased the prevalence of distributed collaboration --- individuals on teams where some or all of the members are geographically distributed and collaboration occurs through technology. 
Since the onset of the COVID-19-imposed work-from-home period, researchers have reported a number of benefits of this collaboration style, including fewer distractions \cite{ford2021tale}, and increased productivity \cite{10.1145/3544548.3580989}. However, the recent increase in this form of collaboration has reaffirmed the challenges that still exist, such as siloing workers \cite{yang2022effects}, impeding creativity \cite{brucks2022virtual}, and reducing awareness of others and their work \cite{yang2022effects, ferguson2022couldn, 10.1145/3544548.3580989}. 

Teams often use meetings to compensate for this lack of awareness. 
Studies of teams transitioning to remote work found that newly scheduled meetings were added weekly when transitioning \cite{ferguson2022couldn}. 
Workers are reporting being in too many unproductive, inefficient meetings, which keep them from doing their own work \cite{perlow2017stop, leeming2020three}. Many meetings can break up a worker's focus time, interrupting their ``flow'', causing task-switching, and fragmenting their attention \cite{Mark2015, laker2022dear}.

\subsection{Focus}


Helping workers focus is a common theme in HCI literature \cite{mark2015focused, mark2017blocking, mark2014bored, saha2023focus}. There is a large body of work looking at how interruptions to focus affect productivity and well-being \cite{blank2020emotional, Brumby2019, mark2018effects, mark2017blocking, mark2016email, mark2015focused, mark2014bored, dabbish2011keep, mark2008cost, 10.1145/633292.633351, czerwinski2004diary, gross2023literature}. Interrupting focus can increase errors, hurt efficiency and productivity \cite{Brumby2019, gross2023literature, 10.1145/633292.633351, 10.1145/3290605.3300845}, and increase sadness and fear \cite{blank2020emotional}. 

Researchers have developed several technical interventions aimed at improving focus in the workplace. For example, 
calendaring software that automatically schedules focus time 
\cite{saha2023focus},
systems which block distracting websites when one is susceptible to interruptions \cite{tseng2019overcoming}, or tools which let others know when someone can be interrupted \cite{10.1145/587078.587125}. 
\citet{10.1145/3290605.3300560} asked whether these distraction-blocking technologies lead to their intended benefit of improved productivity, or if they just redistributed distractions, finding no negative effects of the technology. 

While most past work has taken a technical approach to improving focus, it is both a technical and an organizational problem. Fewer HCI studies have investigated organizational interventions for focus. \citet{mark2016email} found that workers who followed the practice of checking their emails at set intervals were more productive. As work becomes more flexible, organizations are implementing ``core work'' or ``core collaboration'' hours when meetings happen, leaving other time for focus
\cite{Brown_2023}. 

Workplaces, particularly when distributed, are sociotechnical systems where the social or organizational aspects are interdependent with the technical aspects \cite{klein2014we}. Studies of workplaces can be conceived through a sociotechnical lens \cite{10.1145/3544548.3581456, 10.1145/3025453.3025633}, and recent workshops call for the design of sociotechnical supports for workplace collaboration \cite{10.1145/3584931.3611290}. Approaching worker focus through purely organizational interventions, such as meeting-free weeks, may leave gaps or create challenges that technological solutions can help to solve. Similarly, using exclusively technological interventions for focus may result in unintended use of the tools, requiring organizational interventions. In this work, we aim to understand how workers use meeting-free weeks and what gaps exist in their implementation that may be suited for technical solutions. 








\subsection{Collaboration Mechanisms}
 
Modern knowledge work requires collaboration and coordination between teammates \cite{surawski2019knowledge}. 
When collaborating virtually, workers use a number of different technologies: chat platforms, video conferencing software, digital whiteboards, diagramming tools, shared document editors, knowledge management systems, and task management tools \cite{10.1145/3528579.3529171}. In fact, collaborators often use more than one tool at a time, such as having side chats during video calls \cite{10.1145/3411763.3451793}. Two virtual collaboration mechanisms that are widely used and often the focus of HCI studies are audio/video conferencing and instant messaging (IM). 

Video or audio conferencing platforms are the primary way meetings happen between distributed collaborators. By traditional definitions of communication richness \cite{dennis2008media}, video or audio conferencing would be considered a rich communication mechanism as it conveys information in many forms (i.e., through body language and tone). Despite this, this virtual mechanism still has its drawbacks: \citet{10.1145/3462204.3481780} found that online meetings have lower overall experience scores than in-person meetings, and it is harder to coordinate and share ideas.  

Instant messaging technology, sometimes called enterprise social networks \cite{10.1145/1460563.1460674} or enterprise communication platforms \cite{ferguson2023measuring, ferguson2023gender}, are relied on by remote collaborators \cite{ferguson2022couldn}, and are heavily used by the organization studied here. 
Instant messaging platforms are helpful for knowledge and information sharing \cite{zhao2011microblogging},  encouraging informal conversations, enabling faster feedback, promoting awareness \cite{stray2020understanding}, and capturing decision rationale \cite{10.1145/2141512.2141591}. These systems are used for both synchronous and asynchronous collaboration \cite{10.1145/587078.587080}, and have been found to support complex discussions
\cite{10.1145/587078.587081}. However, this specific collaboration mechanism can make it challenging to keep up with multiple threads of conversation, meaning some thoughts go unheard \cite{10.1145/3406865.3418335}, and there are tensions when we extend in-person communication norms to this online setting \cite{10.1145/503376.503410}. 

Many computer-supported cooperative work (CSCW) studies have investigated distributed and remote collaboration 
based on a commonly used tool, such as improving upon video conferencing \cite{tang2023perspectives, 10.1145/3491101.3519792} or instant messaging \cite{10.1145/2998181.2998309, 10.1145/503376.503408}. Other studies investigate collaboration in a technology-agnostic way, such as determining collaborative personas \cite{10.1145/1978942.1979272}. Comparatively fewer papers investigate a collaborative goal in a technology-agnostic way, such as what information people are willing to share, and what they would like their teammates to share \cite{10.1145/3406865.3418308}. In this work, we take a technology-agnostic approach to understand how limiting one common collaboration mechanism influences the use of others toward the collaborative goal of attention negotiation.   

\subsection{Attention Negotiation}
Almost all work requires periods of collaboration and periods of independent work \cite{perlow1999}. In these collaborative periods, a team member may require an answer to a question or feedback on a document; thus, they must direct their colleague's attention to that task at some point in time. At the same time, distributed workers are constantly deciding where to direct their attention. \citet{birnholtz2011framework} describes this as ``joint interpersonal attention'', which requires gathering and displaying signals of availability--- a task that may be straightforward in person but hard online \cite{fogarty2005predicting}. One critical work that we build on is the study of attention negotiation on mobile devices \cite{wohn2015ambient}. 
They found that elements of the technology, the context, and the relationship influenced attention-displaying and gathering behaviour. 

Some methods of attention negotiation are easier than others. Meetings --- or synchronous audio/video conferences --- generally require high levels of attention, although this can vary \cite{10.1145/3313831.3376546}. \citet{10.1145/3411764.3445243} found that workers start splitting their attention between different tasks during larger meetings, and sometimes these low levels of engagement are a deliberate choice \cite{10.1145/3334480.3383080}.

Other collaboration mechanisms have more ambiguous attention requirements, making attention negotiation harder. 
\citet{wohn2015ambient} described how when texting, there is ambiguity in whether the communication was seen, allowing choice in when, and if, users respond to the message at all. Chat-like collaboration platforms do not allow for technical ``start'' and ``ends'' to the needed attention, leading to the feeling of being ``always-on'' \cite{birnholtz2017attending}. For example, \citet{10.1145/2556288.2556973} describe that on instant messaging platforms, people often assume they will receive an immediate reply. Workers sometimes take advantage of this ambiguity, deliberately ignoring requests for their attention in order to protect their focus time \cite{birnholtz2012distance}. \citet{10.1145/3406865.3418571} developed ``private'' statuses that users can show to some people but not others to manage demands for their attention.  

The act of attention negotiation involves gaining an awareness of the availability of your teammates \cite{birnholtz2011framework}.  Awareness has been well-discussed in HCI and CSCW literature \cite{gutwin2004group}, with systems proposed to improve awareness in distributed teams \cite{liu2022exploratory, 10.1145/3406865.3418308}.
Towards effective attention negotiation, \citet{10.1145/3406865.3418308} investigated what workers want to share with their teammates
and found that people were generally eager to share attention information, such as when they were available to be interrupted.
\citet{10.1145/3491102.3517616} further elaborate on tensions that arise when awareness symbols are integrated into messaging platforms, highlighting precisely the challenge in attention negotiation --- needing to protect one's own attentional resources while fulfilling a partner's expectations. We build on this tension by describing workers' strategies to approach this balance.

\subsection{Objectives and Contribution}

In this work, we approach a common problem in distributed knowledge work: a lack of focus. We aim to develop an understanding of how knowledge workers use unstructured time during meeting-free weeks, an organizational intervention to improve focus.
By studying how organizational interventions work, we complement past HCI studies, which have approached focus primarily through a technical lens. Focusing on the organizational intervention allows us to identify existing challenges, such as how workers used other mechanisms to negotiate attention without meetings. 
By investigating both attention-getting and attention-delegating strategies, we hope to tie together work on interruptions and affordances to identify the tensions responsible for these challenges and guidelines to resolve them. We begin by presenting our study of the use of unstructured time during meeting-free weeks, followed by our study of attention-negotiation and compensating mechanisms.

\section{Study 1: Understanding Organizational Interventions}
The first study investigated the ways people use meeting-free weeks and their perceptions of benefits and drawbacks.
\subsection{Organization Background}
This study takes place at an enterprise software development organization comprised of Engineers (majority of employees), Designers, Product Managers, Researchers, and cross-functional teams such as Human Resources and Finance. This organization is predominantly distributed, and most employees work virtually in teams with colleagues located worldwide. Specifically, this organization has teams located in North America, Europe, and Asia. While those located in major cities can go into an office, almost all collaboration happens through technology, primarily Zoom\footnote{\url{https://zoom.us/}}, Google Meet\footnote{\url{https://meet.google.com}}, Gmail\footnote{\url{https://mail.google.com}}/Google Calendar\footnote{\url{https://calendar.google.com}} and Slack\footnote{\url{https://slack.com}}. This organization heavily relies on Slack for remote collaboration; thus, it is frequently referenced in quotes in both studies. This organization exists within the technology sector, which has the greatest percentage of remote workers in the U.S. \cite{wfhreport}.

This organization was chosen for the study as internal employee surveys had identified focus time as a challenge. Leaders implemented meeting-free weeks in June 2021 in order to give workers more time to focus on their core job responsibilities. Each month, leadership encouraged workers to cancel meetings for one pre-determined week --- unless they actively helped make progress on a task. In other words, leaders discouraged meetings but did not ban them; they recognized meetings are necessary for particular kinds of work. Thus, by the time of the study in June-July 2023, workers were settled into the ways they used these weeks. However, internal surveys still suggested mixed benefits and utilization of these weeks, prompting the need to better understand their use. As the company leadership strongly supported these weeks, they supported this research by providing open access to employees across the organization, allowing us to recruit from a broad population and draw conclusions within and across roles.




\subsection{Methods}
For exploratory studies of emerging phenomena, researchers suggest using qualitative methods that allow for flexibility in identifying new themes and ideas \cite{adams2008qualititative}. Since there are few previous investigations of organizations implementing specified meeting-free time \cite{laker2022surprising}, and these provide little description of how workers spend this time, we chose to approach this study qualitatively. We chose semi-structured interviews in order to capture workers' accounts of these weeks and their perceptions of the benefits and drawbacks, allowing us to identify reasons why they may not be broadly beneficial. Based on our research aim of understanding the use of meeting-free time and its connection to focus, we inquired about perceptions of their focus time (whether it improved during meeting-free time), and accounts of how time was spent during meeting-free weeks.
We conducted 17 hour-long semi-structured interviews in June-July 2023. The purpose of the interviews was explained, and participants were given the opportunity to opt out. Participants volunteered for the interview in the course of completing a brief internal employee survey
about focus time. We recruited eight participants who indicated on the survey that meeting-free weeks were not beneficial and nine participants who indicated that they were. 
All participants worked in Product, Design, Engineering, Customer Experience, or the Human Resources department and worked in entirely distributed teams.
The breakdown of the participants is shown in Table \ref{tab:participants}. Engineers make up the majority of the sample, as they make up most of the organization. 

Participants described their most recent meeting-free week (between 2-4 weeks before the interview). We asked them to elaborate on why the week was or was not beneficial and what tasks/activities they performed. To help participants think about how removing distractions influenced focus, we asked them to walk us through their regular week calendars, comparing them to the meeting-free week, identifying cancelled meetings and how time was used instead.
We then inquired about activities performed during meetings and if/how they were completed without the meeting. The interview concluded with perceptions of how meeting-free weeks influenced participants' focus, productivity, creativity, problem-solving abilities, team connection, and general well-being. 

\begin{table}[]
\centering

\caption{Participant information for Study 1. IC = Individual contributor. Benefit group refers to whether they indicated meeting-free weeks were beneficial to them on the related survey (High) or not (Low). Participant numbers represent E=engineer, O=other, I=IC, M=Manager}
\label{tab:participants}
\resizebox{0.45\textwidth}{!}{%
\begin{tabular}{l|l|l|l}
\textbf{ Participant \#} &\textbf{ Role  }              & \textbf{Level }  & \textbf{Benefit Group} \\ \hline
S1EI1                & Engineer            & IC      & Low       \\
S1OI1                 & Designer            & IC      & High    \\ 
S1EM1                 & Engineer            & Manager & High  \\ 
S1OI2                 & Designer            & IC      & High  \\ 
S1EI2                 & Engineer            & IC      & High  \\ 
S1OI3                 & Product Manager     & IC      & Low   \\ 
S1EI3                 & Engineer            & IC      & High  \\ 
S1OI4                 & Customer Experience & IC      & Low   \\ 
S1EI4                 & Engineer            & IC      & Low   \\
S1EM2                & Engineer            & Manager & Low   \\ 
S1OI5                & Product Manager     & IC      & Low   \\
 S1EI5                & Engineer            & IC      & Low  \\
S1EI6                & Engineer            & IC      & High \\ 
S1EM3                & Engineer            & Manager & Low \\ 
S1OM1                & Designer            & Manager & High  \\
S1EM4                & Engineer            & Manager & High  \\ 
S1EI7                & Engineer            & IC      & High \\ \hline
\end{tabular}%
}
\end{table}

 We aimed to identify patterns in how participants across roles and hierarchies used this time, and perceived the intervention. Thus, we used an open-coding method to identify repeating elements. The first author used open coding following the Braun and Clarke thematic analysis method \cite{braun2006using} to analyze interview transcripts. We approached the data set with a lens focused on the use of meeting-free time and the associated outcomes.
The first author organized the identified codes into themes (Step 3 in \cite{braun2006using}), which were then reviewed and further refined (Steps 4 and 5) in discussion with the second author. 
We identified the following high-level themes: 
\begin{enumerate}
    \item Interpretations of meeting-free weeks: interpretations of the organization's encouragement to cancel meetings
    \item Personal characteristics: how individuals' roles and job responsibilities affected their use of unstructured time
    \item Benefits and drawbacks: perceived outcomes of having a meeting-free week
    \item Uses of unstructured time: How tasks were completed without meetings, what participants did instead
    \item Focus time barriers/strategies: individual and interpersonal reasons why workers were or were not able to focus
\end{enumerate}

\subsection{Results}
Our goal was to determine how unstructured time was used during a meeting-free week, and what perceived outcomes resulted.
We performed comparative analysis within and across the themes identified to produce a more concise and actionable conceptual framework that summarizes three observed orientations towards meeting-free weeks (Table \ref{tab:study1}). 


\begin{table*}[]
\centering
\caption{Summary of the observed orientations towards meeting-free weeks.}
\label{tab:study1}
\resizebox{\textwidth}{!}{%
\begin{tabular}{|>{\raggedright\arraybackslash}p{2cm}|%
>{\raggedright\arraybackslash}p{1.5cm}%
|>{\raggedright\arraybackslash}p{4cm}%
|>{\raggedright\arraybackslash}p{5cm}%
|>{\raggedright\arraybackslash}p{4cm}%
|>{\raggedright\arraybackslash}p{4cm}%
|}

\hline
\textbf{Orientation} &
  \textbf{} &
  \textbf{Characteristics} &
  \textbf{Use of unstructured time} &
  \textbf{Benefits} &
  \textbf{Drawbacks} \\ \hline
  \multirow{2}{2cm}{Focus Orientation} &
  Core work &
  Individuals who have many meetings on their calendar during a normal week, but still have individual contributor tasks that need to get done &
  Cancelled almost all meetings to have a lot of unstructured time. Tackled the most complex tasks on their to-do lists.
  These to-do list items reflected core job requirements, not passion projects
  &
  More flexibility in how and when they work, were able to critically think about tasks to produce higher quality work, and increased productivity by clearing important tasks from the backburner&
 Challenges capturing attention of teammates without meetings, felt socially disconnected from their teammates \\ \cline{2-6} 
 &
  Creative work &
  Similar to the core work group, these individuals have many meetings on their calendar, but still have independent tasks to do. Often individual contributors promoted into lead roles
  &
  Cancelled almost all meetings to maximize unstructured time. Spent this time experimenting with new projects, learning something new, or working on things that were exciting/interesting but not a current objective of the team &
  More flexibility in how and when they work, produced high-quality projects, experienced an increase in creativity and well-being &
  From the organization's perspective, they might have been viewed as less productive in the short term, and experienced challenges getting attention from teammates when needed \\ \hline

Collaborative Orientation &
   &
  Typically, people managers or product managers. Often senior in their role, and/or worked in areas where many cases were one-off, so they commonly helped others &
  Spent this time in ad-hoc discussions helping others, in meetings that couldn't be cancelled, interfacing with external teams, protecting their staff from meetings, meeting with hard-to-catch people, having high-level conceptual conversations, or building connections &
  Connecting with hard-to-catch people and having high-level discussions allowed them to do better work, faster, in the following weeks &
  Little to no gain in flexibility in what they work on or how they work, experienced some challenges connecting with people on their team who were focusing, as they are used to immediate replies \\ \hline

Time-Bound Orientation &
   &
  Typically in roles where they have to deal with urgent tasks, or processes that repeat on a regular basis, such as within the engineering department (e.g., on-calls), and particularly leaders in the department &
 Had slightly more unstructured time, but spent it on the same tasks and processes that they normally do. They had no more time to focus on non-immediate to-do list items which they were expected to complete &
  A few hours less of meetings meant regular tasks and processes were completed slightly faster &
   Little to no opportunity to choose what they worked on or when they worked \\ \hline
\end{tabular}%
}
\end{table*}

\subsubsection{Focus Orientation}
When introducing meeting-free weeks, leaders expected workers to use their newfound unstructured time for independent focus on core work, and many did exactly that. Participants exhibiting this orientation typically had a lot of meetings that prevented them from getting their core work done: \textit{
``Where [meeting-free] weeks can be incredibly useful...the stuff you keep having to put off because of the high priority work...being able to circle around and take care of it..''} [S1EI4].

These workers took meeting-free weeks very seriously.
They completely cleared their calendars and turned off notifications to tackle tasks normally interrupted by meetings or coworkers' messages. They saved their large, complex, challenging, or detail-oriented tasks for meeting-free weeks: \textit{``That type of work it's...very detail-oriented...the context switching is very costly...that's what's most beneficial is I know my headiest work, my deepest thinking work I can do there, cause I'm not having to context switch..''}[S1OI2].

These workers reported more flexibility, productivity, and better work product as a result. The increased schedule flexibility allowed them to focus when they had the most energy during the day --- not when they had the fewest meetings. For example, \textit{``During a [meeting-free] week, I can...[work] more like how I prefer, which is to spend the morning...catching up, getting some small tasks completed before deciding...now I'm gonna spend my whole afternoon on this...''} [S1OI1].

With no meetings, this type of worker relied more heavily (or exclusively) on other mechanisms for interdependent tasks (see Study 2). They reported that non-meeting communication was harder to pay attention to, prompted less discussion, and took longer than the same task would take during a meeting. One participant described this, \textit{``And so when we move [standups] to be asynchronous...people will put...`I want to talk about this thing'. But then there's no real forum to pull that out and talk about it...''} [S1OI3]. They also described how they felt less connected to their teammates during this week: \textit{``...it can be a little lonely...because you're not...doing those meetings...you're kind of alone just doing your thing...''} [S1OI1].

While some focused on their core work tasks, others focused on more creative tasks, using unstructured time to take a break from their day-to-day. They learned a new skill or built something for the team that was always pushed to the back burner. As one participant explained, \textit{``The expectation that I've come into [the meeting-free] weeks [with] is that if all I do with that time is learn something new, it was successful...
''} [S1EI6]. Unsaddled by organizational responsibilities, some workers chose projects that reignited their passion for their jobs: \textit{``...[meeting-free] weeks provide a nice time for them to maybe work on more tech debt stuff...it is incredibly useful for engineers to feel like they get to do that kind of technical work that...brought them to the role in the first place.''} [S1EI4]. Other examples of creative work included streamlining team processes, learning a new programming language, building extensions to the product, and rewriting an open-source application.

While this use of focus was not exactly what the organization had intended for meeting-free weeks, participants with this orientation found them extremely beneficial. These activities nurtured their creativity, invigorating them for upcoming weeks when they returned to their core work tasks: \textit{``[meeting-free] week allows you to be more creative. Which...if you're feeding it properly...can breathe life back into the other areas...you're using [this] week to fill back up the tank of creativity that you can then spend on the things that you're doing for work.''} [S1EI6]. While this use of unstructured time was not outwardly ``productive'' to leadership in the short term, workers affirmed it contributed to long-term productivity.

\subsubsection{Collaborative Orientation}

Despite the intention for meeting-free weeks to provide more unstructured time, some participants had no change in the number of meetings, but only in the types of meetings.
People and product managers whose core work happened during meetings wondered if meeting-free weeks even applied to them:
\textit{``I don't want to not have a meeting that is actually gonna unblock us because it's a [meeting-free] week. But then that happens so frequently...you just end up having a bunch of meetings anyways.''} [S1OI3]. 

More senior workers and those who work on a case-by-case basis (e.g., support tickets) frequently helped others in ad-hoc synchronous collaborations which popped up in place of scheduled meetings. As one participant explains,
\textit{``I'm always...supporting my teammates...working through weird tickets...those are just kind of like evergreen. [Meeting-free] week or not, they're always going to happen''} [S1OI4]. Leaders spent time in meetings required to meet deadlines, often to ``protect'' the unstructured time of others on their team. They also attended meetings with people in other parts of the organization that didn't observe meeting-free weeks:
\textit{``The reason that they can go off and spend a lot of time...being inspired...is that the buck stops with me on our team...I will take care of the interruptions...''} [S1EI5]. 

With calendars cleared, other workers opportunistically scheduled meetings with people whose calendars are normally packed, such as executives. As one product manager explains, \textit{``sometimes during [the meeting-free] week, it means the super busy people have less meetings on their calendar...I can put meetings on their calendar.''} [S1OI5]. Participants exhibiting this orientation held high-level conversations about important topics they normally delay, such as long-term planning, roadmaps, strategy, or team cohesion: \textit{``..actually take some time to...clear our eyes a little bit. Look at some of the...review data again...Look at where people are going with their roadmap...It's a really hard thing to fit into like a half hour...''} [S1OM1].





They saved time by calling ad-hoc meetings to work through problems or establish common ground.
However, workers with this orientation did not have the same time and task flexibility that the organization expected from a clear calendar:
\textit{``We can't really stop making progress on the stuff we're trying to make progress on. So continuing those syncs [is] important...to make core decisions...''} [S1EI4].  
 
\subsubsection{Time-Bound Orientation}

The last orientation revolved around urgent or time-bound tasks. Workers in this category included Engineers addressing urgent escalations or managing weekly software deployments. Meeting-free weeks gave them a few hours back, but meetings were never the biggest barrier to focus time: \textit{``Us engineers... we’re obligated to...just in general help...some [operations] tasks we want to do to just help [the product] run smoothly...''} [S1EI1].

Meeting-free weeks had only minor effects on their time (e.g., dealing with an escalation slightly sooner than usual). Other times, nothing would change at all. One participant explained,
\textit{``I was really specialized in the mobile release program...a significant amount of my time is spent doing...recurring activities to support those releases...there is a bunch of things we have to do every week..''} [S1EM2].
Workers with this orientation simply did not have the same time or task flexibility that others did:
\textit{``
They're not...very effective at blocking out a huge amount of time...Because I still got inbound requests, I still have time-sensitive things to have built into my role...''} [S1EI5].

While we note how some roles are more likely to fall into one orientation than others, we also see significant role variation within these orientations.
For example, some senior IC roles (senior designers or engineers) practice the Collaborative orientation, and some managers follow the Focus orientation. Some participants even described how they may use their time in one way for one meeting-free week, but differently in the next. While participants used meeting-free time differently, most reported these weeks were beneficial. As of publication, the organization continues to offer meeting-free weeks (taking into account the improvements in the discussion section).


The results of our first study showed us that during these meeting-free weeks, participants used their time differently, and often differently than other team members. Without knowing that they would have their colleague's attention at the next scheduled meeting, they used what we define as ``compensating mechanisms'' instead --- a suite of other approaches to get the attention of their teammates. While there is typically an understanding that a certain amount of attention is required when participating in a live meeting, this was not always the case when using other mechanisms: \textit{``People just aren't as attentive to [asyncronous messages]..It is much harder to have a discussion in a thread on Slack...than it is in a meeting. And so I feel like people don't respond in those threads when they could''}[S1EI4]. As two out of the three orientations reported struggling to get the attention of their teammates during these meeting-free weeks, we aimed to learn more about how they attempted to get their colleague's attention, and, simultaneously, how others delegated their attention, to understand why this wasn't working.

\section{Study 2: Attention Negotiation and Compensating Mechanisms}
In a second study, we interviewed workers about the challenges of attention negotiation during meeting-free time, to understand what compensating mechanisms they used instead. 

\subsection{Methods}
We recruited from the same departments within the same organization as in Study 1, though there was no overlap in participants between the two studies. Participants volunteered by responding to a message sent to everyone in these departments. Participants worked in distributed teams, collaborating virtually. We also chose to use semi-structured interviews for this study in order to elicit in-depth recounts from our participants, prompting them to share their screens and walk us through their online interactions. Interviews took place during and immediately after a meeting-free week, in the same 2-month period as the first study. All of those who indicated interest completed an interview, for a total of n=11
(Table \ref{tab:participants2}).

\begin{table}[]
\centering
\caption{Participant information for Study 2. IC = Individual contributor.}
\label{tab:participants2}
\resizebox{0.35\textwidth}{!}{%
\begin{tabular}{l|l|l}
\textbf{Participant \#} & \textbf{Role }           & \textbf{Level  } \\ \hline
S2EI1                 & Engineer        & IC      \\ 
S2EI2                 & Engineer        & IC      \\ 
S2EI3                 & Engineer        & IC      \\ 
S2EI4                 & Engineer        & IC      \\ 
S2EI5                 & Engineer        & IC      \\ 
S2OI1                 & Human Resources & IC      \\ 
S2EM1                 & Engineer        & Manager \\ 
S2EM2                 & Engineer        & Manager \\ 
S2EM3                 & Engineer        & Manager \\ 
S2EI6                & Engineer        & IC      \\ 
S2OI1                & Designer        & IC      \\ \hline
\end{tabular}%
}
\end{table}

In these semi-structured interviews, we asked participants to recount instances where they needed someone's attention during a meeting-free week, and when someone needed theirs. Drawing on Birnholtz et al. \cite{birnholtz2011framework}, we centred interviews on the interplay between gathering and displaying awareness information.

Our interview protocol borrowed techniques from narrative inquiry. Narrative inquiry originated from the field of education and comes from the idea that human beings live storied lives 
\cite{connelly1990stories}. Narrative inquiry has been used in the field of human-centred design to understand usability requirements because it elicits contextual information 
\cite{a2016conceptual}.
Other benefits include the ease of eliciting stories, the richness of the data, and the fact that people generally tell the truth in stories \cite{savin2007narrative}. 

Participants walked us through examples of trying to get someone's attention, navigating to the tool and explaining their thought processes. They did the same for examples of others trying to get their attention.
We prompted them to discuss how they noticed the request, whether it was effective in getting their attention, and how they reacted. 

Similarly to Study 1, we had a lens through which we wanted to analyze the interviews based on Study 1 and existing literature: attention negotiation. We used thematic analysis \cite{braun2006using} focusing on identifying codes related to getting or delegating attention. In particular, we sought out examples of attention-getting and attention-delegation strategies and used open coding to identify any other topics. The complete analysis process resulted in the following high-level themes: attention-getting strategies, attention-delegation strategies, things that influence the strategies chosen, and goals of attention negotiation.

\subsection{Results}

\begin{figure*}
    \centering
    
    \includegraphics[width=0.95\textwidth]{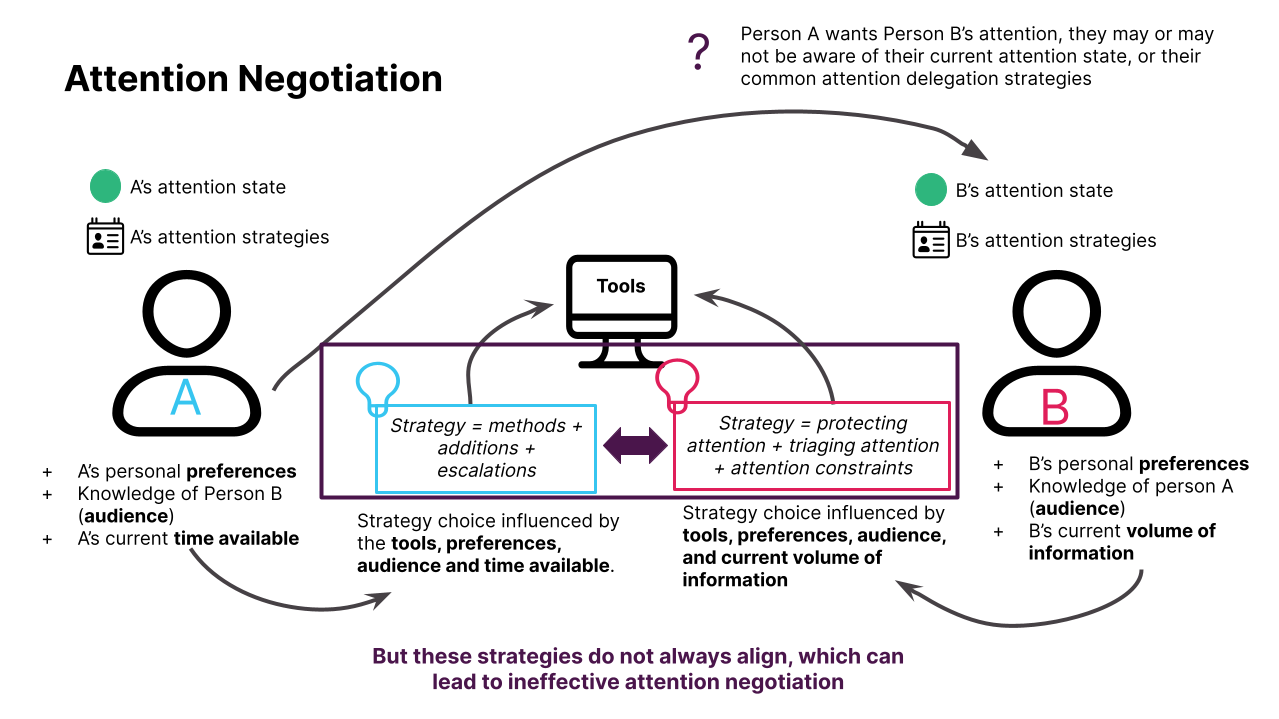}
    \caption{Attention negotiation framework. Each person in attention negotiation has an attention state and attention strategies, which others may or may not be aware of. Person A's attention-getting strategies are comprised of the methods they use, the additions they include in these methods, and their escalation pathways. Their choice of these elements depends on their personal preferences, the way they use their tools, knowledge of the other person, and the time they have available. Person B has attention-delegation strategies comprised of ways of protecting and triaging their attention and constraints on their attention. Their choice of strategy also depends on their preferences, the tools they use, their knowledge of Person A, and the volume of information they are experiencing. If the strategy used to get one's attention does not align with their attention-delegation strategy, 
    tensions arise.}
    \Description{The attention negotiation framework. Two individuals are shown with a computer between them, with arrows moving from the person to the computer, representing interaction through a computer. Above each person is a green dot labelled attention state and a profile icon indicating attention strategies. There is an arrow leading from Person A to Person B's attention state, and question mark demonstrating that person A may not be aware of person B's attention information. Underneath each person are labels for personal preferences, knowledge of the other person, and the current context. Arrows from these labels point to the computer, indicating that these factors influence the attention-getting (Person A) and attention delegation (Person B) strategies used. There is a box around person A's attention-getting strategies and person B's attention delegation strategies, with a double-sided arrow inside and text explaining that if these strategies do not align, tensions result.}
    \label{fig:attention}
\end{figure*}

We developed a framework of attention negotiation that describes how the themes fit together, as shown in Figure \ref{fig:attention}. This simplified framework shows two actors, though attention negotiation can involve more. 
In this example, Person A tries to get Person B's attention using a compensating mechanism (e.g., a Slack message). If Person A's attention-getting strategy does not align with Person B's attention-delegation strategy, then tensions arise. 
Each of these components and their relationship is summarized in Table \ref{tab:attention-framework}.

We describe each component and the tensions that arise when attention-getting and attention-delegation strategies mismatch. 

\subsubsection{Attention-Getting Strategies} \hfill\\
\textbf{Strategies:} 
We found that attention-getting strategies were comprised of a \textbf{communication method} (e.g., direct messages (DMs), team communication (channels), call), \textbf{additions to communication (message elements)} (e.g., indicating urgency), and then \textbf{escalation paths} if the first attempt failed to get their attention. \\

\textbf{Communication types: } Text-based methods included DMs, team-wide communication (e.g., channels), emoji reactions, and changing one's status. After getting attention  through text, participants would
sometimes transition to synchronous audio or video chats. Senders chose DMs for urgent or private matters because they are thought to generate a notification for the recipient: \textit{``DMs are definitely high signal...If I see a DM...especially if it's [from] someone I don't know...it's very attention-grabbing''} [S2EI5]. However, they described preferring channels that do not generate notifications for transparency and status updates: \textit{``
...somebody needs to be able to ask that same question and get the answer that you gave...these things are best done in public form so...people can search for it...''} [S2EM1].

\textbf{Additions:} Participants leveraged features to signal properties about the message. For example, using an @mention 
indicated the message had high relevance to a specific person:
\textit{``
...a mention is basically only for immediate...I need your eyes on this ASAP...''} [S2EM1]. They also signalled urgency using a red circle emoji. CCing or including an executive lent importance to the message: \textit{``I cc'ed them...so that they can go there and see more details if they're interested...they know that it's being addressed...''} [S2EI3]. GIFs and audio clips drew sensory attention in a sea of messages. 

\textbf{Escalations: }When participants didn't get the attention they needed on their first try, they escalated in three ways. Most described that they would use the same mechanism again: \textit{``what a lot of people do is they'll just kind of bump something...you miss a [message]...I might just mention them again...''}[S2EM3]. Others set reminders to try again later, while several found the next best person instead. As one participant described, \textit{``if I feel like I'm not really getting traction with someone, maybe look up another [point of contact]''} [S2EI5]. One participant has a defined escalation path: after following up with a second message, they added ``reactions'' to past messages from that person, hoping to produce a different kind of notification. Failing this, they booked time on the recipient's calendar to review the request: \textit{``the calendar is still sacred for people...I block off time that's free on their calendar. So they have no excuse...''}[S2OI1].

\textbf{Strategy influences: } We identified four main factors that influenced how participants chose a strategy: their \textbf{preferences}, their \textbf{tool usage and understanding of others' usage}, their knowledge of \textbf{the audience}, and \textbf{the time they had available}. 

\textbf{Preferences:} 
They considered whether they preferred synchronous or asynchronous, public or private communication, and how much tolerance they had for ``bothering'' others. One participant described their preference: \textit{``I think I tend to lean towards the [asynchronous] communication part a little heavier than some others on my team...I tend to start drafting a response...''}[S2EI6]. 

\textbf{Tool usage: }They also considered signals from the tools available to them, such as whether a colleague communicated attention information via their status: \textit{``a lot of people don't [change their status]...So it would be nice...to understand how everyone uses [their tools] so that we can at least gauge ourselves for how folks are gonna react''}[S2EI5].

\textbf{Audience: }Characteristics of the person whose attention they were trying to get, such as what they are currently attentive to, their attention-delegation strategies (how they will delegate their attention), and power dynamics influenced the attention-getting strategies used. One participant explained how knowledge of working hours influenced their attention-getting strategy: \textit{``[coworker] is in the [European] working hours...His working hours are pretty different...that's why I was like...he will see this.''}[S2EI3].  

\textbf{Time available: }Lastly, the time that they have available, in that moment, to plan and execute an attention-getting tactic played a role in their choice of strategies. For example, when participants explained that they only had a few minutes before their next meeting, they might send a private DM telling someone to look at a specific conversation instead of trying to find the conversation and mention the person in that conversation to direct their attention.

In summary, knowledge workers' \textbf{preferences, tool usage, audience}, and \textbf{time available} influence the selection of an attention-getting strategy --- itself comprised of \textbf{communication methods, additions}, and \textbf{escalation paths}.

\subsubsection{Attention-Delegation Strategies} \hfill\\
\textbf{Strategies: }
Attention-delegation strategies determine how workers decide to allocate their attention. An individual's attention delegation strategy is comprised of the way they \textbf{protect their focus}, how they \textbf{triage incoming requests for their attention}, and the \textbf{attention constraints} they operate within.

\textbf{Protecting focus: }To protect their focus, workers turned off notifications: \textit{``I usually don’t have the desktop app turned on...I get no notifications''} [S2OI1]. They also ignored inbound requests, assuming critical information would reach them regardless: \textit{``There's a lot of noise...But I only care about this in my box, and then maybe something next to my box...if something [is] really important...my colleagues will...bring it up to me...''} [S2OI1]. Others customized tools to prioritize specific topics or people. 

\textbf{Triaging incoming attention requests: }
When there were many things that needed attention, workers attended to one at a time depending on a rule (e.g., how relevant or urgent the thing is to the current task). One participant explains triaging by channel context: \textit{ ``There are certain channels that I just pay close attention to...We have a...channel...where the engineers put [pull requests]...I pay attention to that channel because it gives me an opportunity to see the changes...''} [S2EM3]. They further explain how at other times they triage based on \textit{``...perceived urgency...if I can tell that this needs a response quickly, I'll just do it right then and there...''} [S2EM3]. Other triage rules included permanence (whether the request would eventually disappear), effort required, specificity (how specific the request was to them) and time (most recent request first). 

\textbf{Attention constraints: }Attention constraints forced attention and prevented the execution of usual attention-delegation strategies (e.g., during ad-hoc synchronous meetings). As described by one participant, \textit{``..it's easier to ignore...or forget about text than doing a...live thing''} [S2EI5]. Workers who always kept communication tools open delegated part of their attention to constantly monitoring new inbound messages: 
\textit{``...if there is something urgent [that comes in], it's very likely I might go and take a look at that while having [a] conversation with you.'' }[S2EM2].

Consider the following example of how these tactics combine into one participant's personal attention delegation strategy: they first skim all of their tools to understand the urgency, then attend to communication directed only at them (e.g., DMs), followed by mentions in public communication. They concluded by reviewing their inbox to triage the remaining requests. They described using the ``inbox view'' to grasp the number of attention requests: \textit{``...when I went into the...meetings, I only had two notifications. Now, I've got seven. Is there anything in there that's urgent? Let me do a quick triage...''} [S2EM3].

\textbf{Strategy influences:} The same factors that influenced the choice of attention-getting strategies also influenced the choice of attention-delegation strategies. These included \textbf{ preferences } (for clearing notifications), \textbf{tool usage} (notification/inbox settings), \textbf{knowledge of audience} (response time expectations, power dynamics) and \textbf{the volume of current attention requests} (the communication context).  

\textbf{Preferences: }Some workers practiced ``inbox zero'' by regularly clearing all notifications before the end of the day, which informed their attention-delegation strategies.
One participant explained, \textit{``I'm kind of like the prototypical inbox zero kind of person who likes to see all of the...unread messages''} [S2EM3].

\textbf{Tool usage: }Some chose their overarching attention-delegation strategy by relying on specific features (e.g., having an inbox view):
\textit{``I'm an [inbox view] user….I have too many channels to just go through and [view each one]''} [S2EI6].

\textbf{Audience:} Triaging strategies were flexible. Power dynamics especially changed how workers delegated their attention. One manager describes that \textit{``it’s like people are waiting to take action or waiting for sort of an approval...So I'm [a] little worried about...what is my factor in terms of their decision time.''} [S2EM2].

\textbf{Volume of information:} Lastly, the volume of requests changes how workers delegate their attention---too many requests meant some were ignored:
\textit{``In this quarter, there's just so much work that it's hard for me. I'm in almost ten channels each day''} [S2EI3].

In summary, one's \textbf{preferences, tool usage, current volume of information experienced}, and \textbf{understanding of the audience }influence the attention-delegation strategies used. These strategies include \textbf{protecting future attention, triaging attention in the present, and operating within constraints}.

\begin{table*}[]
\renewcommand{\arraystretch}{1.5}
\footnotesize
\centering
\caption{Summary and examples of elements of the attention negotiation framework}
\label{tab:attention-framework}
\resizebox{\textwidth}{!}{%
\begin{tabular}{|p{0.1\textwidth}|m{0.4\textwidth}|p{0.1\textwidth}|p{0.25\textwidth}|}
\hline
\textbf{Component} &
  \textbf{Description and Example} &
  \textbf{Subcomponent} &
  \textbf{Subcomponent examples} 
   \\ \cline{1-4}
   \multirow{3}{=}{Attention-getting strategies} &
  \multirow{3}{=}{\justifying{\textbf{Strategy = communication type + communication additions + escalation pathways.} Participants choose a communication type, any additional features to add to the communication, and an escalation pathway in case the first attempt is not successful. For example, participants may choose a personal DM, using an emoji as an urgency indicator, and follow up in a team channel if their DM was not addressed}}  &
  Communication types &
  Individual communication (e.g., DM), ad-hoc synchronous communication, team-wide communication (e.g., channel), emoji reactions 
   \\ \cline{3-4}
  & & 
  Additions to communication &
  Personal call-outs (e.g., @mentions), indicating urgency, context, media, important people   \\ \cline{3-4} & & 
   Escalations &
  Re-sending original communication, personal reminders, finding second best person 
   \\ \cline{1-4} 
\multirow{4}{=}{Attention-getting strategy influences} &
  \multirow{4}{=}{\justifying{The choice of elements within the attention-getting strategy are influenced by \textbf{individual's preferences, their tool usage, their knowledge of the audience, and their current time available}. Participants may choose a DM if they prefer private communication, and ignore others' status indicators if they personally do not use them. They may choose additions such as mentions if they believe the audience will get notified, and not escalate if they have a preference not to be bothersome}} &
  Preferences &
  Synchronous vs. asynchronous, public vs. private, text vs. audio, preference to be bothersome 
   \\ \cline{3-4}
 &
  &
  Tool usage &
  Status features use 
   \\ \cline{3-4}
 &
   &
  Audience &
  Awareness of others' attention state, attention-delegation strategies, power dynamics, target audience 
   \\ \cline{3-4} & & 

   Time available &
  The time one has in the moment to consider their attention-getting strategy 
   \\ \cline{1-4}
  \multirow{3}{=}{Attention-delegation strategies} &
  \multirow{3}{=}{\justifying{\textbf{Strategy = protecting attention + triaging attention + attentional constraints.} Participants aim to protect their attention sometimes, triage their attention when needed, and act within set constraints on their attention. Participants might turn off all notifications for focus time, and triage their attention based on context later, looking first at communication related to their main project. They must act within attentional constraints, such as live meetings}}

   &
  Protecting attention &
  Turning notifications off, ignoring incoming interruptions, prioritizing incoming requests for attention (by source, channel, topic, etc.), reminders 
   \\ \cline{3-4}

 & &  Triaging attention &
  Triaging incoming requests for attention by context, urgency, permanence, effort, specificity, or time \\ \cline{3-4}
 &  & 
  Constraints on attention &
  Constant interruptions, forced attention in meetings 
   \\ \cline{1-4}
  \multirow{4}{=}{Attention-delegating strategy influences} &
  \multirow{4}{=}{\justifying{The choice of attention delegation elements is influenced by one's \textbf{personal preference, how they use their tools, their knowledge of the other person, and their current volume of information}. When one is experiencing many requests for their attention, they may protect their attention by turning off notifications. Their triaging strategies may be influenced by their preferences for certain configurations of the tools they use, such as the order shown in an inbox view. Self-imposed constraints, such as choosing to always have a tool (Slack, email) open on one desktop, constantly consumes some attention}}
  & Preferences &
  Practicing ``zero notifications'' 
   \\ \cline{3-4}
 &
   &
  Tool usage &
  Notification settings, inbox settings 
   \\ \cline{3-4}
 &
   &
  Audience &
  Belief of response time expectations, power dynamics 
   \\ \cline{3-4}

 & & Volume of information &
  Volume of information 
   \\ 
   [7ex] \cline{1-4}
\end{tabular}%
}
\end{table*}

\subsubsection{Tensions in Attention Negotiation}


Despite two years of practicing attention negotiation during meeting-free weeks, we still identified a number of tensions between attention-getting and attention-delegation strategies (Table \ref{tab:tensions}).
Workers used these tensions to describe how more meetings were necessary. To move towards more effective attention negotiation without defaulting to meetings, we need to understand and address these tensions.




According to participants, team-wide channels helped create transparency by consolidating updates in one place. For messages particularly relevant to specific people, participants added mentions:
\textit{``I use mentions...which is like I want to post this for everyone's visibility, but I need action only from two people in that 15-person channel...''} [S2EM2]. However, the volume of communication created more noise than signal. In response, workers used triaging strategies to determine which messages needed their attention, and ignored others. As one participant described, 
\textit{``In terms of people on the team, I am one of the least informed people...I just find it too noisy. I can't keep up with all the announcements...''} [S2EI6].

Tensions also arise when workers do not have information about the attention-delegation strategies of others, so they assume that they are similar to their own. One example of this is how mentions are perceived. While private communication, such as DMs, are broadly believed to be high urgency, mentions in team communication do not have as consistent of an understanding. For some, mentions are only used when something is urgent:
\textit{``mentions...[are] very similar to DMs. It depends on how savvy the person is with configuring their notifications...''} [S2EM1]. However, some teams describe using mentions extremely frequently, so they are not a signal of urgency. One participant described that it depends on where the mention is in the message: \textit{``I feel like I pay more attention when the mention is right at the start of the message as opposed to towards the end...it just makes me think that my attention maybe is not as needed.'' }[S2EI6].

When people thought they had information on other's attention states, they used their preferences to assume incorrect strategies. We found that it was common for individuals to avoid using methods that would send a notification when they knew that someone was away for the day. One participant describes using technical workarounds: \textit{``...if you notice that...their status shows that they're away, I tend to do the classic thing of putting a dash between the @ and their handle so that they don't get the [notification]...''} [S2EM3]. However, others rely on these mentions when they return in order to help triage by what is specific to them. Without these, it is challenging to prioritize items competing for their attention:
\textit{``when you're on vacation, sometimes people will like @ and then dash mention you. So you don't get a notification, which is...annoying...it's kind of nice to get that mention when you get back...''}[S2OI1].

Additionally, participants make assumptions regarding others' attention states, response times, and attention-delegation strategies, when they have little actual information on these. They have to develop these understandings over time. One participant describes their team norms, 
\textit{``DMing people usually gets a quick response, But if you post in a [team] channel, maybe it will take a day''} [S2EI1]. However, in the absence of having this information about the others on their team, participants made assumptions that were not always correct.
While the things like the volume of information (or notifications) a person is dealing with at one time matters when participants are deciding on their attention-delegation strategies, others often don't have this information to consider when trying to get their attention. 

\begin{table}[]
\centering
\footnotesize
\caption{Summary of tensions between attention-getting and attention-delegation strategies and other challenges in attention negotiation.}
\label{tab:tensions}
\resizebox{\columnwidth}{!}{%
\begin{tabular}{|>{\raggedright\arraybackslash}p{2.5cm}|%
>{\raggedright\arraybackslash}p{5cm}%
|}


\hline
\textbf{Tension} &
  \textbf{Description} \\ \hline
Priority of team communication &
  Some put information in team-wide channels with the expectation that everyone will see it, while others ignore it due to high volumes \\ \hline
Assuming attention-delegation strategies of others &
  Some workers only use @mentions in team communication in urgent situations, while others use these so often that they no longer stand out \\ \hline
Over-reliance on limited attention state information &
  Some platforms provided limited views into attention states, such as online vs. offline icons. Workers may over-rely on this information, assuming that if someone is online, they will be attentive to incoming requests
  \\ \hline
\textbf{Challenge} &
  \textbf{Description} \\ \hline
Making assumptions regarding one's attention state or response times &
  Without knowing the attention state of the recipient, workers made assumptions which may or may not be correct \\ \hline
Basing attention-getting strategies on their own preferences &
  Participants based their attention-getting strategies on their own preferences, not the preferences of those they were contacting (often because they did not have that information) \\ \hline
Unawareness of the volume of information of the attention target &
  The volume of information  the target is experiencing influenced their attention-delegation strategies, but we found participants did not have access to this information when choosing attention-getting strategies \\ \hline
\end{tabular}%
}
\end{table}

\section{Discussion}

In this work, we approach the study of workplace focus holistically, recognizing it as a system that requires both organizational and technical considerations. By better understanding how distributed workers adjust to one organizational intervention for focus --- meeting-free time --- we contribute a framework for how workers use compensating mechanisms to continue collaborating. 
Our initial interviews revealed that without the default mechanism of scheduling audio or video meetings, workers faced a major challenge when these compensating mechanisms were insufficient to get their teammates' attention. This motivated a second study, focused on current attention negotiation practices.
We study both sides of attention negotiation, describing tensions between attention-getting and attention-delegation. We build on the body of work discussing the challenges of distributed working styles \cite{10.1145/3544548.3580989} to show that some solutions to these challenges may create new challenges. We show that studying how the organization influences the technical --- or vice versa--- is important to address these trade-offs.  



In the first study, we examined meeting-free weeks in an enterprise software organization to understand how knowledge workers used their newfound unstructured time.
Meeting-free weeks can be successfully implemented to give \textit{some }workers more unstructured time, resulting in more perceived focus time. However, not all workers benefit from these interventions in the same way. Acknowledging this heterogeneity may help organizations design interventions that benefit more workers.

Little work thus far has studied organization-sanctioned meeting-free time, and ours is the first to understand how workers used this newfound time in detail. Similar to \citet{laker2022surprising}, we identified autonomy or time and task flexibility as one of the benefits of this intervention. However, we go further by uncovering how this flexibility is used to derive additional benefits --- such as clearing to-do list backlogs, recharging creativity, or doing the hardest work during their most focused hours of the day.

Generally, organizations are interested in improving focus as a mechanism for increasing productivity. Yet, productivity is notoriously hard to define. We contribute to further describing perceptions of the relationship between focus and productivity in knowledge workers --- we identify ways in which meeting-free weeks lead to increases in perceived productivity. Similarly to \citet{10.1145/3290605.3300845}, we describe how when participants had fewer meeting interruptions, they felt that they were able to do more work, and complete it to a higher quality. On the other hand, we show how meeting-free time made it harder to get the attention of others, which could end up causing a bottleneck in interdependent teams.
We also build on a more holistic view of how time is spent, as suggested by \citet{10.1145/3313831.3376586}, in our Focus - Creative work orientation. While this might not have been viewed as a ``productive'' use of time by the organization, participants perceived long-lasting benefits that resulted from using unstructured time for creative work. We contribute to the discussion of how and when focus might lead to some forms of productivity. 


Further, we found that roles explained some patterns in orientation, though not all. For example, people managers frequently exhibited Collaborative or Time-Bound orientations with less unstructured time. 
Past work found that managers view interruptions as part of their job \cite{hudson2002d} because these interruptions allow them to deal with problems earlier. Yet, in studying a firm-wide shift to distributed work, \citet{yang2022effects} found that managers experienced a larger increase in requests for their attention than individual contributors. Managers might need these focus-time interventions most, even if they don't see them as part of their role.
More research is needed to test how new interventions, role-specific resources or training might amplify the benefits of unstructured time for all. 

Our participants also reported feeling disconnected from colleagues during meeting-free weeks.
Synchronous meetings are the only ``face-to-face'' time that distributed workers have, whereas in-person workers are likely to see their colleagues in the hallway or lunch room, helping to maintain awareness and connection during meeting-free weeks. Understanding the impact of missing context for short periods of time (a week at a time) and how to fill this in asynchronously is an important area of future work. 


In our second study, we view attention negotiation as a multi-actor process that affects workers' focus. During meeting-free weeks, workers need to get the attention of their colleagues without their default daily or weekly meetings, balancing 
the need to protect their own attention while addressing others' many requests for it \cite{10.1145/3491102.3517616}. We build on existing studies of collaboration mechanisms \cite{mcfarlane2002scope} by focusing on all parties in the collaborative goal of attention negotiation, identifying the 
tensions and challenges resulting from strategies that are misaligned. We show that these mechanisms are interconnected --- people choose their attention-getting strategies based on the mechanisms their audience uses or reacts to.



This work brings together coordination mechanisms defined in past HCI work to show their interaction. For example, we describe how ``escalation pathways'' from \citet{birnholtz2011framework} are influenced by personal preferences and knowledge of the audience. We expand on the tensions that arise from relying (or not) on status indicators to estimate response times \cite{o2014everyday}.
We connect elements of \citet{mcfarlane2002scope}'s taxonomy of interruptions by describing how one's characteristics (preferences) and communication method influence how successful an interruption is at getting attention. We found that deliberately ignoring requests is an attention protection strategy, similar to \citet{LEE2023102983} and \citet{mcfarlane2002scope}, and by focusing on the act of negotiation, we show that this can result in unique escalation attempts. 
Lastly, we elaborate on the tension between protecting one's own attention resources and fulfilling their colleague's expectations, as outlined in \citet{10.1145/3491102.3517616}, by describing how workers currently try to balance these needs, and how we could design tools to better support them. 



Our attention negotiation framework also ties together theories of tool affordances. By studying both sides of attention negotiation, we describe tensions which prevent some of these affordances from being realized. For example, immediacy, an affordance of IM platforms \cite{10.1145/358916.358975}, is derived from assumptions of how others delegate attention, and we found that not everyone prioritized, or even used, IM notifications.
Similarly, \citet{Anders_2016}
highlights how 
being able to customize your notifications on Slack affords compartmentalization of projects/tasks. On the other hand, we found that because notifications are so customizable, everyone sets them differently --- yet assumes everyone has them set similarly.

\subsection{Organizational Implications}


We begin by sharing implications for organizations looking to provide workers with more unstructured time. First, meeting-free time can improve perceptions of focus. However, not all workers benefit equally; personal characteristics, roles, and task types make a difference. Organizations can start with talking to workers across teams and hierarchies to understand the types of meetings they're in, and considering the following recommendations based on the orientations we described in our study: 

\begin{itemize}
    \item \emph{Focus orientation.} Encourage all departments to hold simultaneous meeting-free weeks so cross-functional workers can focus. 
    Set clear expectations about goals for the week --- whether progressing core work tasks, or reinvigorating creativity through exploration and learning, perhaps offering time for both. The challenge is in identifying weeks suitable for all teams; in large organizations, there are likely to be important collaborative periods for some on any given week. Organizations can work to avoid known seasons, like road mapping/strategic planning, and alternate meeting-free weeks to ensure that all parts of the organization can benefit.
    \item \emph{Collaborative orientation.} Let workers know that a meeting-free week can still include collaboration. Encourage scheduled meetings to be cancelled, with ad-hoc collaborations used only when needed, ensuring that only the required people are pulled in.  Limit these to particular days to maximize uninterrupted time for others who prefer to focus.
    \item \emph{Time-bound orientation.} Investigate how workers currently manage unavoidable interruptions and find ways to minimize them. 
    Designate the smallest group needed to handle emerging incidents so others may avoid these interruptions and benefit from unstructured time. This group should be appropriately managed to not overwhelm or overburden one group of individuals, having urgent work tasks bleed into their personal time. Managers should monitor the employees in these groups, and design a dedicated process for calling in extra help when needed.
\end{itemize}

Many tensions from our second study could be improved with organizational or team norms for how to collaborate without meetings. 

Setting norms or expectations for how to communicate is not uncommon in workplaces. For example, in-office teams develop norms for when it is appropriate to interrupt someone (likely not when they are mid-conversation), when you should be available for collaboration (often sometime between 10am-3pm), and appropriate response times for emails or phone calls. Distributed teams can and are \cite{Brown_2023} developing similar norms around important ways to communicate. Norms do not need to drastically change how workers interact with their tools. We see the following areas as instances where norms may help to ease some of the tensions identified.

\emph{Team agreements.} Teams can create an explicit set of expectations for communication, including tools/channels to monitor or ignore, response time expectations, days/times for collaborating or focusing, appropriate times to turn off notifications, and what features should be used (e.g., setting an away status). This should not be created by one team member (e.g., manager) and forced on all; it should be a collaborative discussion that takes into consideration what information is critical to whom, who might be experiencing information overload, preferences, and evolving best practices \cite{Kupp_2023}. Teams would likely have to experiment with different agreements.

 We provide an example of the successful implementation of team agreements from one of our participants. As manager of the team, our participant started an agreement for channel monitoring and let his team members have control over the channels they agreed to.  
If they agreed to monitor a channel, it was expected that communication within that channel would be read in one hour. This agreement helped team members delegate their attention effectively; they knew which channels they had to prioritize, and which they could ignore altogether. It also helped our participant to know where to put information that he wanted everyone to see.

\emph{Reconsider the need for organization-wide communication.} We found that one of the reasons distributed workers report feeling overwhelmed by requests for their attention
was due to the number of broadcast communications they received. Consider the internal company communications sent to the entire company each day, compounded with regular updates from the pillar, department, and team. Participants described that the same updates were often cross-posted in more than one of these broadcasts. These broadcasts were the ones deemed least important and more likely to be ignored, yet they cluttered their inboxes, making attending to critical information challenging. 
Organizations may need to reconsider what information is shared broadly and how often; perhaps instead, they should post to internal wikis that workers can search if needed, or managers can be asked to relay messages to their teams as they see fit.

\subsection{Design Implications}
We now revisit our study of attention negotiation to draw out
design implications for the tensions and challenges identified in Table \ref{tab:tensions}.

Design changes can address tensions with the priority of team communication. Some expect team-wide messages to be seen by the whole team, but too many messages harm the signal-to-noise ratio. One potential solution could be to allow senders to choose when messages are sent. While scheduled send already exists, senders need to set the time for each message, which might not be convenient for everyone. A bulk update, sent regularly at times specified by each individual, may allow all team members to stay up to date without creating noise during the workday. Past work found that this type of batching for emails increased productivity \cite{mark2016email}.
Message senders could mark each message as ``send now'' or ``send in update'' depending on the urgency (even having the default be ``send in update'', and only having to change this when urgent), while any users @mentioned would be automatically notified as usual, allowing users to choose who is updated immediately. Workers could set their personal preferences for when to receive batched updates for these less urgent matters, such as their lunch hour or the end of their workday, making them less disruptive. This relies on the sender's assumption of who this information is most important for, and could prevent one from adding to a discussion if they were notified too late. Thus, options to visit certain channels and see new information (while not receiving notifications/visual cues for them) must still be available.
This could result in large volumes of information to sort through at one point in the day, particularly for those involved in multiple highly communicative teams. In these instances, automated summarization of these updates could also be used, and enterprise communication tools have recently announced the development of these capabilities\footnote{\href{https://slack.com/blog/productivity/product-innovations-dreamforce-2023}{https://slack.com/blog/productivity/product-innovations-dreamforce-2023}}.

Features can also address the tension that arises when a worker assumes the attention-delegation strategy of their coworkers. Some participants saw mentions as urgently directed towards them, while others did not. 
Additional properties of mentions may help recipients prioritize (e.g., @bob vs. @bob-urgent). Tools could display these in different categories, order, or colour --- or they could be used to override ``do not disturb'' settings \cite{fogarty2005predicting}. However, this feature must be paired with organizational norms for how to use it --- if everything is marked as urgent, nothing really is.
This feature could be accompanied by a pop-up ``are you sure you want to send this message'' confirmation, even describing the use case in the particular organization (e.g, ``urgent mentions should only be used for business-critical purposes''). 

The rest of the tensions and challenges can be addressed with design features that provide more information about the current state or the working style of the attention target, which is particularly important for building awareness in distributed work \cite{birnholtz2011framework, hancock2009butler, wohn2015ambient}. Tensions arise when limited signals in tools cause workers to incorrectly infer others' attention states, or 
when they are open to being contacted \cite{10.1145/503376.503410}. While basic availability features exist on some collaboration systems, such as Microsoft Teams \footnote{\href{https://learn.microsoft.com/en-us/microsoftteams/presence-admins}{https://learn.microsoft.com/en-us/microsoftteams/presence-admins}}, the online/offline dichotomy may be too coarse. Simply knowing someone is online does not guarantee they would be attentive to incoming requests, as they may have their notifications off or are triaging to only pay attention to certain projects. More detailed availability signals can help people plan their attention-getting strategies, and we see two suggestions for this.  

One is for status indicators to include more detailed attention states, such as what users are attentive to currently and when they are available for interruptions. These could be pre-set options that are easy to select, such as ``focusing'', ``available for collaboration'', or ``working towards a deadline'', which would reduce the burden of creating custom statuses but still require users to set these themselves. \citet{10.1145/3406865.3418308} found that workers want to disclose when they are interruptable to their teammates, and teammates want this information available on the platforms they use. Alternatively, some statuses can be automated. For example, features that display when one is experiencing a higher-than-normal volume of attention requests,  or when one has a slower-than-normal response time (indicating meetings or focused work), may help to set response time expectations. 

Systems might also display users' attention-delegation strategies to their coworkers, such as a ``how to contact me'' section on a profile. This could 
include statements like ``I prioritize my team channels during the workday (\#engteam1-)'' or ``please only use DMs for urgent matters, direct everything else to...'' .
This information could be shared in automated replies, like out-of-office emails, that direct team members to other channels, or set response time expectations.

These features come at a cost to privacy, requiring a balance between providing justification for statuses and protecting privacy \cite{10.1145/3491102.3517616}.
There should always be an option to set a default ``away'' status that communicates unavailability to teammates 
without providing additional details. Workers may still be hesitant to disclose the projects they are focusing on if they don't align with strategic priorities, and may feel pressured to be ``available'' if others are. As with any technology, developing an explicit understanding of appropriate use within the team is critical. 



Finally, workers may differ in how they prefer to be contacted depending on the person, topic, or task. The methods that are most effective at getting one's attention (e.g., a DM) may not be the same as the methods most effective for completing the task once attention has been successfully negotiated (e.g., debugging by sharing screens in a video call). Thus, we should design tools to ease the transition between methods, such as allowing users who are engaging in a channel discussion to jump into a call, or using DMs to catch one's attention and direct them to a team-wide discussion happening elsewhere. While it may be ideal to develop tools that are effective at drawing attention anywhere, one of the only widely agreed-upon strategies from our study was that DMs are highly attention-grabbing. Designing tools that make it easy to transition from an attention-getting mechanism to the most appropriate communication mechanism for the task may improve attention negotiation.

\subsection{Future Work}

Our early exploration of meeting-free time in an enterprise organization leads to additional questions. We studied a primarily distributed organization where workers lack the informal communication co-located workers have \cite{ferguson2022couldn}, and some researchers believe that many elements differentiate in-person from distributed work \cite{olson2000distance}. Future work could explore meeting-free time and the resulting attention negotiation when workers are co-located. 


Our interview methods did not permit us to formally evaluate the design of the intervention. With some evidence that there is a relationship between the two \cite{laker2022surprising}, future work might study this topic with different methodologies. An experimental design would allow causal claims about the impact of length and frequency of meeting-free time on outcomes like focus and productivity.
We caution researchers and organizations from relying exclusively on objective measures, as things like productivity are notoriously hard to define and measure \cite{10.1145/3313831.3376586, 10.1145/3290605.3300845}, and there are many subjective outcomes of meeting-free time that we identified, such as improvement in well-being.


Lastly, we call for future work to design and build features or plug-in tools for virtual collaboration platforms 
that implement the design implications described above. We believe that the features that could make the most impact on attention negotiation are awareness-building features specifically aimed to communicate attention states and attention-delegation strategies to one's team members. 



\section{Limitations}
Because we studied one organization with three years of experience with meeting-free weeks, 
we are limited in the generalizability of our conclusions. We were able to identify details of the different orientations workers have towards meeting-free time, but other organizational characteristics like size, industry, and tools may impact these results. Meeting-free time is a relatively recent intervention; organizations trying it for the first time will differ from the one we studied, where workers were familiar with the practices and tools.
Methodologically, interviews helped us understand this relatively new phenomenon \cite{bell2002narrative},
 but prevent us from studying the topic at scale (e.g., through surveys, diary studies, or calendar logs). Interviews with 28 workers represent a small portion of the entire organization, and we cannot claim that we identified every orientation or attention negotiation practice. 

\section{Conclusion}

In response to concerns about the number of meetings workers have, organizations have started implementing meeting-free time with the goal of improving focus. Yet focus is generally approached using technical interventions. Our work provides a preliminary understanding of this organizational intervention, and the accompanying challenges that result, through two complementary interview studies with distributed knowledge workers. We discovered three main orientations that workers had toward unstructured time: Focus, Collaborative, and Time-Bound. Each orientation has benefits and drawbacks, and differences in these orientations resulted in reliance on compensating mechanisms to negotiate attention. The second interview study aimed to understand attention negotiation. 
The resulting framework 
illuminates the tensions and challenges that exist when attention-getting strategies and attention-delegation strategies misalign. Approaching focus time in distributed workplaces holistically, we identify how organizational interventions may be beneficial, but create new challenges, which may be candidates for technical solutions.  

\begin{acks}
We would like to thank the Research and Analytics team at Slack for their feedback and support while conducting this study. We thank Dr. Jeremy Birnholtz for his thoughts and insights into the framing of this study, and Dr. Alison Olechowski and Paula Aoyagui for their reviews and edits. We also thank the CHI reviewers and chairs for their detailed and constructive comments.  
\end{acks}

\end{document}